\documentclass{elsart}
\usepackage{graphicx,natbib,amssymb}
\journal{New Astronomy}

\def\I{\rm {\scriptsize I}}
\def\astrobj#1{#1}
\def\bibcode#1{}

\begin{document}
\begin{frontmatter}

\title{The SED in the hot continuum of the symbiotic binary 
       \astrobj{AR\,Pavonis}}
\subtitle{I. Tests with current models}

\author{A. Skopal\thanksref{fn1}}

\thanks[fn1]{E-mail: skopal@ta3.sk}

\address{Astronomical Institute, Slovak Academy of Sciences, 
         059\,60 Tatransk\'{a} Lomnica, Slovakia }

\begin{abstract}
We present the spectral energy distribution (SED) in the continuum of 
the eclipsing symbiotic binary \astrobj{AR\,Pav} between 0.12 and 
3.4\,$\mu$m. 
This revealed a high luminosity of the hot object in the binary, 
$L_{\rm h} \sim 2\,200\,(d/4.9\,{\rm kpc})^{2}\,L_{\odot}$. 
We introduce a method of disentangling the total continuum spectrum 
into its individual components of radiation for current models 
of symbiotic binaries. 
Applying a standard ionization model we show that the configuration 
of \astrobj{AR\,Pav} differs significantly from that typical for symbiotic 
binaries during their quiescent phases.
The best fit of the observed SED is provided by radiation of 
a simple blackbody accretion disk with 
$L_{\rm AD} \sim 1\,700\,(d/4.9\,{\rm kpc})^{2}L_{\odot}$, 
which is embedded in an extended hot corona with 
$T_{\rm e} = 40\,000\,\pm$5\,000\,K and 
$L_{\rm N}\sim 500\,(d/4.9\,{\rm kpc})^{2}L_{\odot}$. 
This basic configuration of the hot object explains also the observed 
wavelength-dependent depth and width of the eclipse profile. 
The standard thin disk model requires a high accretion rate 
$\dot M_{\rm acc} \gtrsim 2 \times 10^{-4}\,M_{\odot}\,yr^{-1}$
onto the central star with a radius $R_{\rm acc} \gtrsim 2\,R_{\odot}$ 
to balance the observed luminosity. 
Even irrespectively to the disk model, the accretion process limits 
$R_{\rm acc} > 0.1\,R_{\odot}$ for 
$\dot M_{\rm acc} > 1.7 \times 10^{-5}\,M_{\odot}\,yr^{-1}$ 
and
$M_{\rm acc} = 0.75 - 1.0\,M_{\odot}$, 
which precludes a white dwarf to be the accreting star.
Application of models with the disk and the boundary layer shows that 
the far-UV spectrum is not consistent with a large amount of a hot 
radiation from the boundary layer. However, the presence of such 
a boundary layer in the system is indirectly indicated through 
the strong nebular emission. This solution suggests that 
the hotter inner parts of the disk including the boundary layer 
are occulted by the disk material in the direction to the observer. 
\end{abstract}
\end{frontmatter}

\section{Introduction}

\astrobj{AR\,Pavonis} was discovered by \citet{mayall} as an eclipsing 
P-Cygni type star with a period of 605 days. 
\citet{andrews} found that the eclipsed object is highly variable 
in both brightness and size. On the basis of ultraviolet spectroscopy 
\citet{hutchings83} confirmed this view. 
\citet{bruch} observed a periodic ($\sim P_{\rm orb}$) wave-like 
variation, which occurred in the visual light curve between eclipses 
from 1985.7. As the most satisfying explanation they suggested 
a brightness modulation due to variation in the mass transfer rate 
from the red giant to the hot component. 
Analysing historical 1889--1998 light curve, 
\citet{sk00a} identified the real change of the orbital 
period at a rate $\dot P = -3.5\pm 0.8\,\times\,10^{-5}$ between 
1896 April and 1985 September. They also noted that during quiescence, the
light curve has characteristic features similar to those observed in CVs. 
Recently, \citet{sk01a} refined the mean orbital period 
to 604.45$\pm 0.02$ days. 

As to the \astrobj{AR\,Pav} configuration, \citet{thh74} proposed 
a binary model, 
in which the red giant fills its Roche lobe and loses mass to an evolved, 
$\sim$30\,000\,K hot, compact object surrounded by a moderately dense 
gas cloud or a thick ring. 
\citet{kw84} found that the observed UV continuum
colours for \astrobj{AR\,Pav} are well comparable to those derived from 
a model of an accretion disk around a main sequence accretor. 
Recently, \citet{schild01}, having new radial velocities 
of the red giant, determined its mass function to 0.055\,$M_{\odot}$ and 
the projected rotation velocity of 11$\pm 2\,\rm km\,s^{-1}$. Assuming 
co-rotation, they derived the giant radius of 130\,$R_{\odot}$ and by 
comparing its adequate stellar parameters with RGB/AGB evolutionary tracs 
in the HR diagram, they suggested the giant mass as 2\,$M_{\odot}$, 
and, consequently from the mass function, the hot component mass 
as 0.75\,$M_{\odot}$. Based on these parameters, \citet{schild01} 
suggested that the hot component in \astrobj{AR\,Pav} is most probably 
a white dwarf and the cool component is deeply inside of its Roche lobe 
(however, as we will show later in this paper, the giant in this 
system does most probably fill its Roche lobe). Most recently, 
\citet{q02}, having radial velocities of both components, 
determined their masses as 
$M_{\rm g} = 2.5\,M_{\odot}$, $M_{\rm h} = 1\,M_{\odot}$. They obtained 
the radial velocity curve of the hot star under assumption that the 
broad emission wings of the H$\alpha$, H$\beta$ and He\,\I\I\,4686 
profiles originate from the inner accretion disk or an envelope around 
the central star. In addition, the authors suggested that the central 
absorption in hydrogen line profiles is formed in the neutral part of 
the giant's wind, which is concentrated in the orbital plane. 

In this paper we analyse the SED in the continuum between 
0.12 and 3.4\,$\mu$m with the aim to determine a basic configuration 
of the hot object in \astrobj{AR\,Pav}. In Sect. 3 we introduce individual 
components of radiation often used in interpretation of the spectra 
of symbiotic stars. In Sect. 4 we compare the observed SED and 
those given by current models of symbiotic stars, including 
a standard ionization model and models with an accretion disk. 
Finally, in Sect. 5 we discuss their physical plausibility and 
suggest basic structure of the hot object. 

\section{Observations} 

To reconstruct the SED of \astrobj{AR\,Pav} in the range 
of 0.12 to 3.4\,$\mu$m 
we used the broad-band optical and infrared $UBVRI$ and $JHKL$ 
photometry and the IUE (International Ultraviolet Explorer) low-resolution 
spectroscopy. 

The optical/IR magnitudes were summarized by \citet{sk00a} and some 
new observations were recently published by \citet{sk00b}. 
To determine the star's brightness in eclipses we used photometric 
measurements of \citet{andrews} and \citet{mal82}. 
Observations in the $R_{\rm C}$ and $I_{\rm C}$ bands of the Cousins 
system were transformed into the Johnson system according to 
\citet{bessell}. Stellar magnitudes were converted to fluxes 
according to the calibration of \citet{lena99}. 

The ultraviolet fluxes were taken from Fig. 5 of 
\citet{schild01}. We used average values from those 
measured out of eclipses (i.e. between orbital phases 
$0.91 > \varphi > 1.1$; the rms within 10-15\%) 
and values from the mid eclipse. To get better coverage of 
the ultraviolet continuum we added a few values from Fig. 2 of 
\citet{hutchings83} and measurements presented in 
Table 4 of \citet{kw84}. 

Finally, we dereddened observations with $E_{\rm B-V}$ = 0.26 
\citep{kw84} using the extinction curve 
of \citet{mathis} in the UV/optical region, while in the 
IR wavelengths we used reddening ratios of \citet{sm79}. 
Table 1 summarizes the observations used in our paper. 

\section{Spectral energy distribution}

Figure 1 shows the 0.12 -- 3.4\,$\mu$m SED in the continuum 
of \astrobj{AR\,Pav}. 
This suggests a high luminosity of the hot object, similar to that of 
its cool companion. 
The out-of-eclipse continuum is characterized by: (i) A round far-UV 
shape with decreasing fluxes to shorter wavelengths, and (ii) a flat 
profile in the near-UV/optical region. Both the steep far-UV 
($\lambda < 1\,500$\,\AA) continuum and the prominent Balmer jump 
in emission -- characteristic features of the continuum of symbiotic 
stars during quiescence -- are not present. 
We will fit the out-of-eclipse continuum by that including an accretion 
disk, but also test whether a standard ionization model of symbiotic 
binaries can explain these characteristics. 

During eclipses, the ultraviolet continuum weakened by a factor of 
$\sim$3--5, it is flatter than that observed out of eclipses and 
displays a discontinuity at $U/B$. This suggests the nebular-type 
spectrum of a very hot gas. We will fit the uneclipsed continuum 
by the nebular radiation of a high electron temperature. 
In the following sections we describe simple models of the considered 
components of radiation, which we use to fit the observed SED. 
%
%
\begin{figure}[t]
\centering
 \includegraphics[angle=-90,width=9cm]{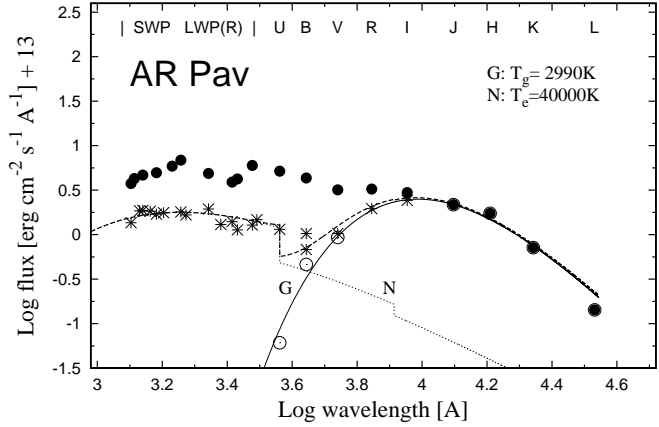}
 \caption[ ]{
Dereddened fluxes of the \astrobj{AR\,Pav} continuum given by 
the low-resolution 
IUE spectra, optical $UBVRI$ and infrared $JHKL$ photometry (Table 1). 
Measurements made during eclipses are marked by ($\ast$), out of eclipses 
by ($\bullet$), and open circles ($\odot$) correspond to fluxes of 
the giant derived from its spectral type \citep{mal82}. 
Observations during eclipses can be approximated by nebular continuum of 
$T_{\rm e} = 40\,000$\,K (N) and fluxes of the giant by a Planck 
function of the colour temperature $T_{\rm g} = 2\,990$\,K (G). 
The dashed thick line represents their sum. 
} 
\end{figure}
%
%
\begin{table} 
\begin{center}
\caption{Continuum fluxes between 0.12 and 3.4\,$\mu$m.}
\begin{tabular}{rccccc}
\hline
$\lambda$[\AA] & $F_{\rm obs}$ & $F_{\rm der}$ 
               & $F_{\rm obs}$ & $F_{\rm der}$ & Ref. \\
& \multicolumn{2}{c}{in eclipses} & \multicolumn{2}{c}{out of eclipses} & \\
\hline
1270       & 0.15 & 1.36 & 0.40 & 3.75 & 3 \\
1300       &      &      & 0.48 & 4.27 & 4 \\
1350       & 0.23 & 1.85 &      &      & 3 \\
1380       & 0.25 & 1.87 & 0.64 & 4.67 & 5 \\
1455       & 0.26 & 1.83 &      &      & 3 \\
1520       & 0.25 & 1.70 & 0.73 & 4.96 & 5 \\
1600       & 0.27 & 1.76 &      &      & 3 \\
1700       &      &      & 0.91 & 5.88 & 4 \\
1808       & 0.28 & 1.80 & 1.07 & 6.87 & 5 \\
1875       & 0.26 & 1.67 &      &      & 3 \\
2200       & 0.20 & 1.94 & 0.50 & 4.88 & 3,4 \\
2400       & 0.23 & 1.30 &      &      & 3 \\
2600       & 0.29 & 1.40 & 0.80 & 3.89 & 3,4 \\
2700       & 0.25 & 1.13 & 0.93 & 4.22 & 5 \\
3000       & 0.34 & 1.28 & 1.59 & 6.00 & 5 \\
3100       & 0.40 & 1.47 &      &      & 3 \\
3645       & 0.52 & 1.60 & 1.71 & 5.16 & 2 \\
4400       & 0.41 & 1.03 & 1.66 & 4.33 & 2 \\
4400       & 0.27 & 0.68 &      &      & 6 \\
5500       & 0.49 & 1.03 & 1.52 & 3.19 & 2 \\
7000       & 1.14 & 1.99 & 1.86 & 3.25 & 2 \\
9000       &      &      & 1.86 & 2.66 & 2 \\ 
12500      &      &      & 1.77 & 2.17 & 1 \\
16200      &      &      & 1.50 & 1.74 & 1 \\
22000      &      &      & 0.65 & 0.72 & 1 \\
34000      &      &      & 0.14 & 0.14 & 1 \\
\hline
\end{tabular}
\end{center}
Fluxes are in $10^{-13}\,\rm erg\,cm^{-2}\,s^{-1}\,\AA^{-1}$ \\
Ref.: 1 - \citet{gw73}, 
      2 - \citet{mal82}, 
      3 - \citet{hutchings83}, 
      4 - \citet{kw84},
      5 - \citet{schild01},
      6 - \citet{andrews}
\end{table}

\subsection{Nebular continuum} 

In the conditions of ionized gaseous nebulae in symbiotic stars 
(a dense medium with $n_{\rm e} \sim 10^{6} - 10^{11}\,\rm cm^{-3}$), 
the main contributor to the near-UV/optical continuum is the free-bound 
(f-b) emission from hydrogen. Less important are the f-b emission 
of He\,\I\I\ and He\,\I. 
Due to the dominance of the H\,\I\ f-b emission and its basically 
similar profile with that of He\,\I\ and He\,\I\I\ 
\citep[see Fig. 1 of][]{bm70}, and also for the sake of simplicity, we 
calculate here only H\,\I\ f-b and f-f nebular contributions. 
   The nebular flux is proportional to $\int \!n_{+}n_{\rm e}\,dV$, 
where $n_{+}$ and $n_{\rm e}$ is the concentration of protons and 
electrons, respectively. Accordingly, the observed flux, 
$F^{\rm obs}_{\lambda}$ ($\rm erg\,cm^{-2}\,s^{-1}\,\AA^{-1}$), 
of the nebular continuum at the wavelength $\lambda$, can be 
written as 
\begin{equation}
  4\pi d^{2} F^{\rm obs}_{\lambda} = \varepsilon_{\lambda}\!\int_{V}\!
                                      n_{+}n_{\rm e}\,dV,
\end{equation}
in which $d$ is the distance to the object, $V$ is the volume of the 
ionized zone and $\varepsilon_{\lambda}$ ($\rm erg\,cm^{3}\,s^{-1}\,
\AA^{-1}$) is the total (f-b + f-f) volume emission coefficient 
per electron and per ion in the wavelength scale. We calculated 
them according to expressions given by \citet{bm70} adopting 
the Gaunt factors to be unity. In spite of this approximation, our 
procedure reproduces the tabulated values \citep{bm70,gurzadyan} 
within $\sim$5\% at the blue side of the Balmer discontinuity 
and within 1\% or less at other wavelengths for electron temperatures 
$T_{\rm e}\,\sim$10\,000 -- 20\,000\,K. The difference is smaller for 
higher $T_{\rm e}$. 

\subsection{Blackbody accretion disk and the boundary layer continuum}

Because of having defined the continuum only by fluxes at selected 
wavelengths (i.e. only the continuous radiation can be investigated), 
we will assume a steady-state, time-independent accretion disk around 
a central star of mass $M_{\rm acc}$ and radius $R_{\rm acc}$. 
Further, the accretion disk is optically thick and radiates locally 
like a blackbody \citep[e.g.][]{tylenda}. The source of its energy is 
the gravitational potential energy of the accretor. The observed flux 
distribution of the disk, $F_{\lambda}$, is then given by contributions 
of blackbody annuli integrated through the entire disk. 
The flux emitted by each annulus at the radial distance $r$ 
is weighted by its area, 2$\pi\,r\,dr$. Thus the observed flux 
distribution of a disk at a distance $d$ can be written as 
\begin{equation}
 F_{\lambda} = \frac{2\pi\cos(i)}{d^{2}}\int_{R_{\rm acc}}^{R_{\rm d}}\!
               B_{\lambda}(T_{\rm eff}(r))\,r\,{\rm d}r,
\end{equation}
where $i$ is the inclination of the disk, $R_{\rm d}$ its radius, and 
the radial temperature structure of the disk, 
\begin{equation}
T_{\rm eff}(r) = T_{\ast} \Big[\frac{R_{\rm acc}}{r}\Big]^{3/4}
             \Big\{1 - \Big[\frac{R_{\rm acc}}{r}\Big]^{1/2}\Big\}^{1/4}, 
\end{equation}
where 
\begin{equation}
 T_{\ast} = 4.10\times 10^{4}R_{9}^{-3/4}({\rm acc})\,
            M_{1}^{1/4}({\rm acc})\,
            \dot M_{16}^{1/4}({\rm acc}) ~ \rm K ,
\end{equation}
in which $R_{9}$ and $M_{\rm acc}$ are the radius and mass of the 
accretor and $\dot M_{16}$ is the accretion rate \citep{w95}. 
The notation $X_{n}$ means that the parameter $X$ is expressed as 
fraction of $10^{n}$ units (the units are $cm,g,s$). 
A useful characteristic of the disk is its maximum temperature, 
which occurs at $r = (49/36)R_{\rm acc}$ and has a value 
0.488\,$T_{\ast}$. 

The final stage of the mass transferred throughout the accretion disk 
is its deceleration and landing onto the accreting star. 
This happens in the so called {\em boundary layer} (BL), in which 
the inner disk material loses its kinetic energy to match the surface 
velocity of the star. If the accreting star rotates at a small fraction 
of its breakup velocity, then the BL emits a luminosity comparable 
to that of the disk. Contrary, when the star rotates close to breakup, 
there should be little emission from the BL \citep[e.g.][]{w95}.
The geometry of the BL can be approximated by a ring with 
thickness $2H$ surrounding the accreting star. Then the luminosity, 
$L_{\rm BL}$, of its area, $2\pi R_{\rm acc} 2H$, for a non-rotating 
star is comparable to that of the accretion disk. So, 
\begin{equation}
 L_{\rm BL} = 4\pi R_{\rm acc} H \sigma T_{\rm BL}^4 \equiv 
  \frac{1}{2}\,\frac{G\,M_{\rm acc}\,\dot M_{\rm acc}}{R_{\rm acc}} = 
 L_{\rm AD}. 
\end{equation}
The effective temperature of the BL can be derived with the help 
of Eq. 5, and according to \citet{w95} it can be approximated as 
\begin{equation}
 T_{\rm BL} \approx 
             2.9\times 10^5 M_{1}^{1/3}({\rm acc})\,
                            R_{9}^{-7/9}({\rm acc})\,
                            \dot M_{18}^{2/9}({\rm acc}) ~ \rm K. 
\end{equation}
In the following section we use these components of radiation 
to fit the observed SED. 
%
%
\begin{figure}[t]
\centering
 \includegraphics[angle=-90,width=9cm]{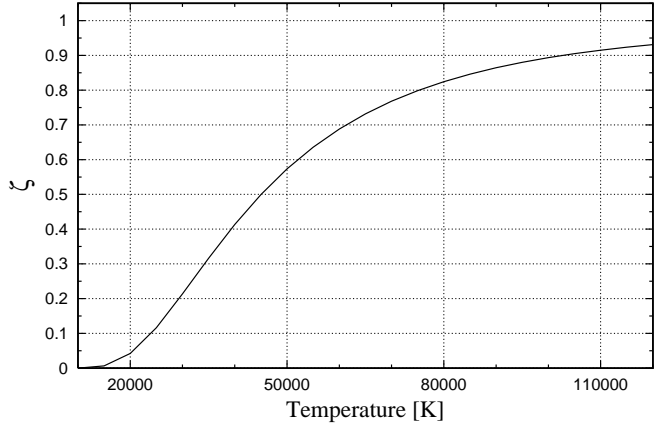}
 \caption[ ]{
Parameter $\zeta$ as a function of the blackbody temperature. 
It represents a fraction of the hot star luminosity below 912\,\AA\ 
(Sect. 4.2).
} 
\end{figure}
%
%
\begin{figure}[t]
\centering
 \includegraphics[angle=-90,width=9cm]{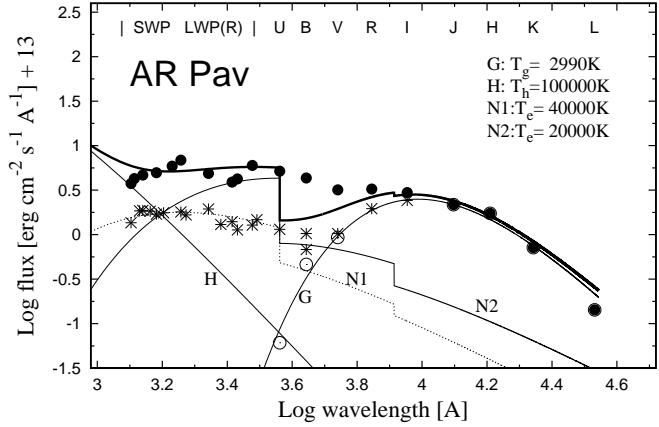}
 \caption[ ]{Comparison of the observed \astrobj{AR\,Pav} continuum 
and calculated 
according to an ionization model: Hot star ($T_{\rm h} = 10^{5}$\,K) 
ionizing the circumbinary medium giving rise to the nebular continuum 
($T_{\rm e} = 20\,000$\,K). Solid thick line represents superposition 
of all components (G+H+N1+N2). 
These models are not able to give satisfactory agreement with 
observations (Sect. 4.2). 
} 
\end{figure}

\section{Modeling the UV/optical continuum}

\subsection{The uneclipsed nebula} 

Figure 1 shows a faint component of the nebular radiation seen during 
eclipses. We compared a grid of models of the nebular continuum to 
these fluxes and selected that with $T_{\rm e} = 40\,000 \pm 5\,000$\,K, 
which corresponds to the best fit to observations (Fig. 1). 
Luminosity of this nebular emission is 
$L_{\rm N}\sim 500\,(d/4.9\,{\rm kpc})^{2}L_{\odot}$. 
The uncertainty in $T_{\rm e}$ results from that in fluxes as given by 
the IUE observations (about 10\%, in average). An analysis of the 
$\chi^{2}(T_{\rm e}$) function is described in \citet{sk01}. 

\subsection{SED of an ionization model}

During quiescent phases of symbiotic binaries the hot component 
($T_{\rm h} \approx 10^5$\,K) ionizes a fraction of the neutral 
wind of the cool giant giving rise to the nebular continuum 
($T_{\rm e} \approx 1.5 - 2.0 \times 10^4$\,K). As a result 
the UV/optical continuum is given by superposition of the stellar 
and nebular components of radiation. 
Below we test this possibility for the case of \astrobj{AR\,Pav}. 

The maximum luminosity of a blackbody hot star of a radius $R_{\rm h}$ 
and temperature $T_{\rm h}$, which can be converted into the nebular 
radiation is 
\begin{equation}
  4\pi R_{\rm h}^{2}\int_0^{912}\!
  \pi B_{\lambda}(T_{\rm h})\,{\rm d}\lambda = 
                                \zeta\,L_{\rm h}  ~~~\rm erg\,s^{-1},
\end{equation}
where the factor $\zeta\,(< 1)$ represents the fraction of the hot 
star luminosity, $L_{\rm h}$, below 912\,\AA. It is a strong function 
of the blackbody temperature between 20\,000 and 60\,000\,K (Fig. 2). 
Under assumption that the ionized medium is optically thick in 
the Lyman continuum, i.e. all the photons with $\lambda <$ 912\,\AA\ are 
converted into the nebular emission, the maximum nebular luminosity can 
be expressed as 
\begin{equation}
L_{\rm neb}^{\rm max} = \int_{912}^{\infty}\!\varepsilon_{\lambda}\,
                     {\rm d}\lambda\int_V \!n_{+}n_{\rm e}\,{\rm d}V = 
                     \zeta L_{\rm h} ~~~\rm erg\,s^{-1}. 
\end{equation}
Thus if the near-UV continuum is of the nebular nature, then for 
the observed flux, for example at 
$\lambda$=3\,600\,\AA, 
$F^{\rm obs}_{3600} \sim 5\times 10^{-13}\,\rm erg\,cm^{-2}\,s^{-1}\,\AA^{-1}$ 
(Table 1), the distance 
$d$ = 4.9\,kpc 
and the emission coefficient 
$\varepsilon_{3600} = 2.6\times 10^{-28}\,\rm erg\,cm^{3}\,s^{-1}\,
\AA^{-1}~(T_{\rm e}=20\,000$\,K), 
we get by using Eq. 1 the emission measure 
$EM = \int_{V}\!\! n_{+}n_{\rm e}\,dV  = 5.5\times 10^{60}\,\rm cm^{-3}$ 
and according to Eq. 8 
\begin{equation}
L_{\rm neb}^{\rm max} = EM\!
          \int_{912}^{\infty}\!\varepsilon_{\lambda}\,{\rm d}\lambda\, 
          \dot{=}\, 1\,230\,L_{\odot}. 
\end{equation}
The corresponding minimum of the hot star luminosity, which is capable 
of producing such nebular emission, is 
$L_{\rm h}^{\rm min} = L_{\rm neb}^{\rm max}/\zeta$ = 1380\,$L_{\odot}$ 
for $T_{\rm h} = 10^{5}$\,K ($\zeta$ = 0.89). 
To compare the model with observations, first, we scaled 
the emission coefficient $\varepsilon_{\lambda}$ to fluxes 
between $\lambda$3000--3600\,\AA, where the nebular 
component of radiation dominates the SED in the ionization model. 
The scaling factor 
$k_{\rm N} = F^{\rm obs}_{\lambda}/\varepsilon_{\lambda} = EM/4\pi d^{2}$. 
Second, we scaled the hot star radiation with a factor 
$k_{\rm h} \geq L^{\rm min}_{\rm h}/4\pi d^{2}\sigma T_{\rm h}^{4}$
to match observations in the far-UV. 
Figure 3 shows our best solution, given by superposition of the 
nebular continuum with $T_{\rm e}$ = 20\,000\,K and a hot stellar
source ($L_{\rm h} \sim 2.5\times L_{\rm h}^{\rm min}$, 
$T_{\rm h} = 10^{5}$\,K), which roughly matches the observed 
UV continuum with a larger deviation at the far-UV. 
Cases with $T_{\rm e} > $ 20\,000\,K significantly 
exceed the far-UV fluxes, and those with $T_{\rm e} < $20\,000\,K 
cannot produce its rather flat profile. All cases do not match 
the optical continuum. 
Finally, the hot star with a lower temperature transforms 
a smaller fraction of its stellar radiation into the nebular 
(e.g. $T_{\rm h} = 5\,10^{4}$\,K gives $\zeta$ = 0.57), which then 
requires too high $L_{\rm h}^{\rm min}$ other than to match 
the SED in the far-UV. 

This analysis shows that the observed SED of the \astrobj{AR\,Pav} 
UV continuum 
does not correspond to that of the ionization model of symbiotic 
binaries during quiescent phases. 
On the other hand, the presence of highly ionized emission lines 
in the UV spectrum \citep{hutchings83} and the nebular 
emission during eclipses indicate a source of ionizing photons 
in the system. 

\subsection{SEDs with accretion disk}

In accordance with previous suggestions, here we will fit the 
UV/optical/near-IR continuum of the hot object in \astrobj{AR\,Pav} by 
models with an accretion disk as introduced in Sect. 3.2. 

\subsubsection{SED with a single accretion disk}

Here we calculated a grid of models of fluxes given by Eq. 2 
for fixed values of $M_{\rm acc}$ and $\dot M_{\rm acc}$, but 
different $R_{\rm acc}$, and selected that giving the best fit to 
the observed fluxes under conditions of the least square method. 

According to \citet{schild01}
we adopted the mass of the donor (the giant star) 
$M_{\rm g} = 2\,M_{\odot}$ 
and that of the accretor 
$M_{\rm acc} = 0.75\,M_{\odot}$. 
The rate of period change 
$\dot P = -3.5\times 10^{-5}$ 
\citep{sk00a}, the mass ratio 
$q = M_{\rm g}/M_{\rm acc} = 2.7$ 
and the orbital period of 605 days yield the accretion rate 
$\dot M_{\rm acc} = 1.7\,10^{-5}\,M_{\odot}\,yr^{-1}$ for the Roche 
lobe overflow if the transferred mass and its angular momentum are 
conserved in the system \citep[][see Sect. 5.1 for more details]{pw85}. 
Using these values of 
$M_{\rm acc}$ and $\dot M_{\rm acc}$ to fit the component of 
radiation supposed to be due to the accretion 
disk (= SED - N - G; Fig. 4) by Eq. 2, we found the best solution 
for $R_{\rm acc} = 0.78\,R_{\odot}$ (Fig. 4, Table 2). 
We integrated the disk contributions to its outer radius 
$R_{\rm d} = 10\,R_{\odot}$, which can be considered as the smallest 
dimension of the disk for these parameters. This is because the disk 
temperature between the surface of the accretor and $r \sim 10\,R_{\odot}$ 
is higher than 5\,000\,K, and thus radiative contributions of 
this part of the disk are important to explain the UV/optical SED 
(i.e. $R_{\rm d}$ cannot be smaller than $10\,R_{\odot}$), 
but the disk annuli with $r > 10\,R_{\odot}$ (i.e. $T(r) < 5\,000$\,K) 
do not contribute significant emission at UV/optical wavelengths, 
and are negligible with respect to the radiation of the giant 
in the IR region (i.e. $R_{\rm d}$ can be larger than $\sim 10\,R_{\odot}$). 
Superposition of the considered components of radiation here 
(G, N, AD) matches well the observed SED of \astrobj{AR\,Pav} (Fig. 4). 
This model expresses the round profile of the far-UV continuum, 
the flat near-UV/optical continuum and a relatively significant 
(with respect to the ionization model) contribution to the near-IR 
continuum of the giant. Note that we observe about 0.2\,mag deep 
eclipse even in the $I$ band \citep{mal82}. 

The accretion disk luminosity here, 
$L_{\rm AD} \sim 1\,700\,(d/4.9\,{\rm kpc})^{2}L_{\odot}$, 
can be balanced by the energy liberated in 
an accretion process for large accretion rates of 
$\dot M_{\rm acc} > \,10^{-4}\,M_{\odot}\,yr^{-1}$ (Table 2). 
In addition, for the orbital inclination of $\approx 70^\circ$ 
(Sect. 5.2), and assuming the disk plane to be coplanar with it, 
the total disk luminosity is by a factor of about 1.5 larger 
($L_{\rm AD}^{\rm obs} \approx 2\cos(i) L_{\rm AD}$). As discussed 
below (Sects. 5.1 and 5.2), observations do not contradict 
such a large $\dot M_{\rm acc}$ for a certain period of evolution 
of the giant star in \astrobj{AR\,Pav}. 
%
%
\begin{figure}[t]
\centering
 \includegraphics[angle=-90,width=9cm]{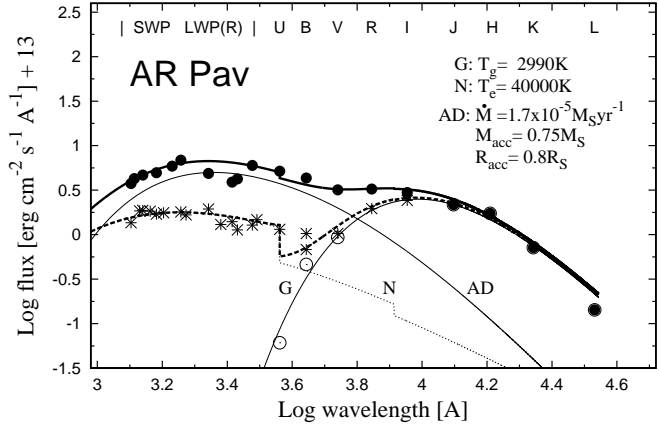}
 \caption[ ]{
Comparison of the \astrobj{AR\,Pav} continuum and the calculated flux 
distribution for an accretion disk (AD). Solid thick line represents 
the resulting continuum given by the sum of all components of 
radiation (G+N+AD). Dashed thick line as in Fig. 1. This model 
is described in Sect. 4.3.1. 
}
\end{figure}

\subsubsection{SED with the boundary layer and accretion disk}

To calculate the flux contributions from the BL we assumed that this 
region can be approximated by a blackbody at a constant temperature 
$T_{\rm BL}$ (Eq. 6). In our procedure we fitted superposition of 
fluxes from the accretion disk and the BL, 
\begin{equation}
 F_{\lambda} =  F_{\lambda}(AD) + \pi  B_{\lambda}(T_{\rm BL}), 
\end{equation}
to the observed SED, where $F_{\lambda}(AD)$ is given by Eq. 2. 
First, we scaled contributions from the BL with a factor 
\begin{equation}
 k_{\rm BL} = \int_{\lambda}\!\! F_{\lambda}(AD)\,{\rm d}\lambda\,/\,
 \sigma T_{\rm BL}^4 
\end{equation}
to satisfy conditions of Eq. 5. Second, we calculated a grid of models 
for 
$M_{\rm acc}$ = 0.75 and 1.0\,$M_{\odot}$, 
$R_{\rm d}$ = 30\,$R_{\odot}$, 
$\dot M_{\rm acc}$ 
between 
$1.0\,10^{-5}$ and $5.0\,10^{-4}\,M_{\odot}\,yr^{-1}$ 
and 
$R_{\rm acc}$ between 0.1 and 3.4\,$R_{\odot}$. 
Finally, we selected models, for which the function 
\begin{equation}
\chi^{2}(M_{\rm acc},\dot M_{\rm acc},R_{\rm acc}) = \Sigma 
 (F_{\lambda}^{\rm obs} - 
 F_{\lambda}(M_{\rm acc},\dot M_{\rm acc},R_{\rm acc}))^{2}/N 
\end{equation}
reached a minimum. We used fluxes from 1270 to 7000\,\AA\ (Table 1), 
but omitted those at 2200, 2600 and 2700\,\AA, i.e. N = 11. 
The function $\chi$ is plotted in Fig. 5. As can be seen from 
this figure there is no satisfactory solution. The best fits 
correspond to radii $R_{\rm acc}$ between 0.7 and 2.5 $R_{\odot}$ 
and the BL temperature $T_{\rm BL} \sim $ 40\,000 -- 60\,000\,K. 
The impossibility to obtain a better solution is reflected by 
the too low required temperature of the BL, of which the flux 
dominates the far-UV wavelengths. Figure 6 shows an example. 
The results of this section suggest 
that the BL emission is not important to fit the \astrobj{AR\,Pav} SED. 
Below, in Sect. 5.3, we discuss the BL emission in more detail. 
%
%
\begin{figure}[t]
\centering
 \includegraphics[angle=-90,width=9cm]{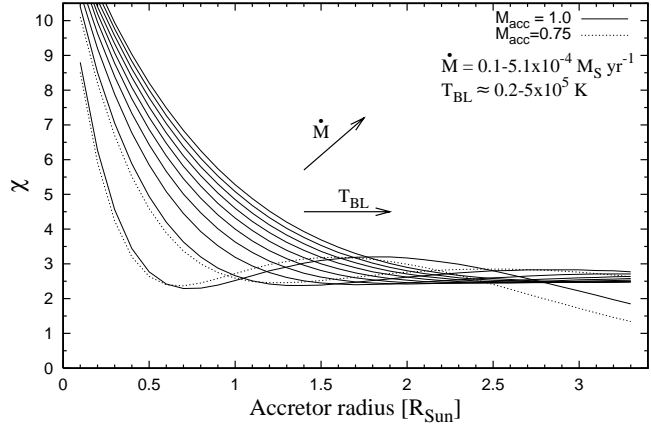}
 \caption[ ]{
The quantity $\chi$ as given by Eq. 12 in modeling the observed 
SED by the BL and accretion disk contributions. 
Solid lines represent models for $M_{\rm acc} = 1\,M_{\odot}$ 
and $\dot M_{\rm acc} = 0.1 - 5.1\times 10^{-4} M_{\odot}\,yr^{-1}$. 
Compared are also two $\chi$ functions for $M_{\rm acc} = 
0.75\,M_{\odot}$ (dotted lines). The minima of $\chi$ correspond 
to $T_{\rm BL} \sim$\,40\,000 - 60\,000\,K. Their large values 
reflect poor fits of these models to observations 
(Sect. 4.3.2, Fig. 6).
}
\end{figure}
%
%
\begin{figure}[t]
\centering
 \includegraphics[angle=-90,width=9cm]{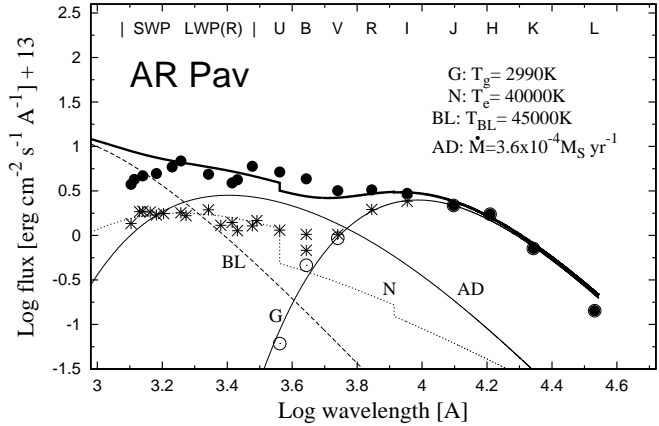}
 \caption[ ]{
One example from the best solutions of fitting the SED by 
the accretion disk and its BL flux contributions. The fits are 
very poor, because of a significant contribution of the BL in 
the far-UV wavelengths. This model is described in Sect. 4.3.2
}
\end{figure}
%
%
\begin{table} 
\begin{center}
\caption{Examples of models of the SED for selected values of 
         $\dot M_{\rm acc}$.}
\begin{tabular}{lccccc}
\hline
$R_{\rm acc}$ & $R_{\rm d}$ & $\dot M_{\rm acc}$ & $L_{\rm acc}$ & 
$T_{\rm max}/T_{\rm BL}$ & $\chi/10^{-13}$ \\
$[R_{\odot}]$ & $[R_{\odot}]$ & $[M_{\odot}\,yr^{-1}]$ & 
$[L_{\odot}]$ & [K] &  \\
\hline
\multicolumn{6}{c}{Accretion disk (Sect. 4.3.1)} \\
   3.7   & 40    & $ 1.7\,10^{-3}$  &  5410  &16600/46100&   0.08 \\
   1.7   & 20    & $ 1.7\,10^{-4}$  &  1180  &16000/50600&   0.09 \\
   0.78  & 10    & $ 1.7\,10^{-5}$  &   257  &16800/55600&   0.11 \\
   0.01  &~~0.1  & $ 3.3\,10^{-11}$ &  0.04  &16500/88500&   0.07 \\
\multicolumn{6}{c}{Accretion disk and the boundary layer (Sect. 4.3.2)} \\
   2.0   & 30    & $ 2.1\,10^{-4}$  &  1240  &15600/46700&   2.51 \\
   2.5   & 30    & $ 3.6\,10^{-4}$  &  1700  &15100/44300&   2.53 \\
\hline
\end{tabular}
\end{center}
Notes: 
In all cases $M_{\rm acc} \equiv 0.75\,M_{\odot}$, 
$L_{\rm acc}=1/2\,G\,M_{\rm acc}\,\dot M_{\rm acc}/R_{\rm acc}$. \\
%
%
\end{table}

\section{Discussion}

\subsection{A mass transfer problem in \astrobj{AR\,Pav}}

The accretion rate of $1.7\,10^{-5}\,M_{\odot}\,yr^{-1}$ (Sect. 4.3.1) 
derived from the orbital period change corresponds to the conservative 
mass transfer from the giant star onto its companion. 
However, successful fits of the UV continuum require at least one order 
of magnitude higher rates to balance the observed luminosity. 
This discrepancy can be caused by mass loss from the system, which 
makes the orbital period increase \citep[cf.][]{sob}. 
So, the mass transfer rate and the mass loss rate rival each other 
in the orbital period change. 
Mass loss from \astrobj{AR\,Pav} is indicated by radial velocities 
of absorption components of hydrogen lines. Already \citet{mayall} 
revealed a P-Cygni type of hydrogen line profiles on three 
plates from 1900, 1901 and 1908. \citet{thh74} measured absorption 
components of the H$\gamma$ and/or H$\delta$ line profiles between 
1953 and 1972. They determined an average position of these absorptions 
at -85.2\,km\,s$^{-1}$. Recently, \citet{q02} measured radial velocity 
of the central absorption component of H$\alpha$ at -96\,km\,s$^{-1}$ 
on spectra taken between 1988 and 2001. These values are shifted by 
-17 to -28\,km\,s$^{-1}$ relative to the systemic velocity 
\citep[-68\,km\,s$^{-1}$,][]{schild01,q02} and thus reflect 
a mass loss from the system. 
  Under such conditions the observed decrease in the orbital period 
requires a much higher mass transfer than that given by conservative 
case. Therefore, the mass transfer rate (i.e. the accretion rate) 
of $1.7\,10^{-5}\,M_{\odot}\,yr^{-1}$ can be considered only as 
a lower limit. This fact allow us to estimate the accretor radius 
$R_{\rm acc} > 0.1\,R_{\odot}$ for 
$M_{\rm acc} = 0.75 - 1.0\,M_{\odot}$ 
to balance the observed accretion luminosity irrespectively to 
the accretion disk model. Such the large $R_{\rm acc}$ precludes 
a white dwarf to be the accreting star in \astrobj{AR\,Pav}. 

\subsection{Roche lobe overflow or a wind accretion?}

Observations and the model energy distributions discussed in this paper 
suggest that the sole energy source in \astrobj{AR\,Pav} is accretion. 
This view 
is also supported by the behaviour in its historical, 1889-2001, 
light curve, in which no large (5 - 7\,mag) outbursts as in symbiotic 
novae have been observed \citep[see][]{sk01a}. 
In addition, the light curve has characteristic features similar to 
those observed in CVs \citep{mayall,sk00a}, which suggests that 
the basic principles of mass transfer in \astrobj{AR\,Pav} 
are similar to those of CVs. These arguments suggest that the red giant 
in \astrobj{AR\,Pav} fills up (or is close to) its Roche lobe. 
Contrary, if the giant in \astrobj{AR\,Pav} underfills its Roche lobe, then 
the mass transfer is via accretion from a stellar wind. However, 
this cannot be true. If this were the case, there would have been 
a huge (unrealistic) mass loss via the wind 
(at least of $10^{-3}\,M_{\odot}\,yr^{-1}$) to get the accretion rate 
required to balance the observed luminosity of the hot object. 
In that case, the wind mass loss would make the orbital period 
{\em increase}, while the observations show that the period 
{\em decreases}. The latter can only be due to the Roche lobe overflow. 

Other observational results allow this possibility. For example, 
the giant's radius inferred from eclipses, 
$R_{\rm g}/A = 0.30\pm 0.02$ \citep{sk00a}, is equal to its 
volume-equivalent Roche radius for the orbital inclination of 
70$^{\circ}$ \citep{q02}. The V-shape of the eclipse profile supports 
this possibility \citep[e.g. Fig. 8 of][]{sk00a}. Further, the 106-day 
periodic variation with the amplitude of a few $\times$0.01\,mag 
observed in the visual light curve \citep{sk00a} could be understood 
as a response of the giant's radius to its mass loss. Generally, 
after losing an amount of mass, the giant will restore its 
hydrostatic equilibrium, which leads to variation in difference 
of the radius of the star and its Roche lobe. In the case 
of the presence of a deep convective envelope in the giant star, 
its radius rapidly increases and the star can run into a violent 
mass loss instability \citep[see][for more detail]{sob, slh}. Such 
a transient high mass transfer rate from the red giant to the hot 
component could be responsible for a higher level of the \astrobj{AR\,Pav} 
activity observed during 1900-01, 1935-36 and 1985-99 \citep{sk00b}. 

\subsection{Boundary layer emission in \astrobj{AR\,Pav}}

The nebular component of radiation indicates the  presence of 
a hot ionizing source in \astrobj{AR\,Pav}. In the models with 
accretion disk the ionizing source is represented by the BL. 

The nebular emission has a luminosity 
$L_{\rm N}\gtrsim 500\,(d/4.9\,{\rm kpc})^{2}L_{\odot}$ 
(= uneclipsed nebula without emission lines), which places strong
constraint on the BL luminosity and ionizing capacity. 
For parameters of the fit shown in Fig. 4 and described in Sect. 4.3.1,
$L_{\rm BL} (\equiv L_{\rm acc}) \sim 1200\,L_{\odot}$ and
$T_{\rm BL} \sim  51\,000$\,K (Table 2). To balance the $L_{\rm N}$ 
luminosity, a fraction of the BL emission below 912\,\AA\ has to be 
\begin{equation}
  L_{\rm BL}(\lambda < 912\,\AA) =
  \zeta\,L_{\rm BL}\, >\, 500\,L_{\odot}.
\end{equation}
For $L_{\rm BL} = 1200\,(d/4.9\,{\rm kpc})^{2}L_{\odot}$,
Eq. 13 yields the factor $\zeta > 0.42$, which forces
$T_{\rm BL} > 40\,000$\,K (Fig. 2). 
Thus the BL characterized by these quantities can balance well
the observed nebular luminosity, $L_{\rm N}$. 

However, our modeling the SED {\em with} the BL (Sect. 4.3.2, Table 2) 
does not provide satisfactory fit (Figs. 5, 6). 
This implies that we cannot see the hotter inner parts of the disk
and that any radiation generated in the BL is absorbed and diffused
in the disk in the direction to the observer, but it is free in
directions to the poles to ionize there the circumstellar matter. 
This situation thus produces two components of radiation of a very 
different temperature regimes observed in the ultraviolet spectrum: 
A relatively cool continuum, which was investigated in this paper, 
and a hot ionizing source seen through the nebular emission. 

It is plausible that the lack of the BL emission in the continuum
results only from a different real geometry of the accretion disk
(probably more extended at its edge) than we adopted in our modeling
(Eq. 2). So, if the disk is rather thick in its outer parts, then we
can see essentially only its outer edge, which occults the BL at
the disk's center for a high orbital inclination. Such a disk 
structure was suggested for the first time by \citet{ph94} for 
interacting binaries in which the accreting star in non-degenerate. 

\subsection{Configuration of the hot object and the eclipse profile}

Our models of the observed SED (Sects. 4.1, 4.3.1) suggest that 
the hot object in \astrobj{AR\,Pav} consists of an accretion disk around 
a central star embedded in a hot corona. This basic 
configuration can be independently verified by the eclipse 
profile, which is a function of wavelength. In the following two 
sections we therefore explore its two main characteristics -- the 
different eclipse depth and width in the UV/optical/near-IR region 
on the basis of our model with accretion disk. 

\subsubsection{Depth of eclipses}

Already \citet{hutchings83} noted that there is no
total eclipse below 3600\,\AA, and 70\% of the hottest continuum
(below 1400\,\AA) is seen even at central eclipse. Recently
\citet{schild01} demonstrated this behaviour
on the ultraviolet IUE light curves.

Generally, the eclipse depth is given by the ratio of fluxes observed
{\em in} and {\em out} of eclipse. In view of our models, it is 
determined by the ratio of the N+G to all considered components 
of radiation. In Fig. 7 we compared the modeled eclipse depth 
according to the solution shown in Fig. 4 and that observed 
in the light curves \citep[cf.][]{schild01,andrews,mal82}. We can 
see that our model explains well the wavelength-dependent depth 
of eclipses. 
%
%
\begin{figure}
\centering
 \includegraphics[angle=-90,width=9cm]{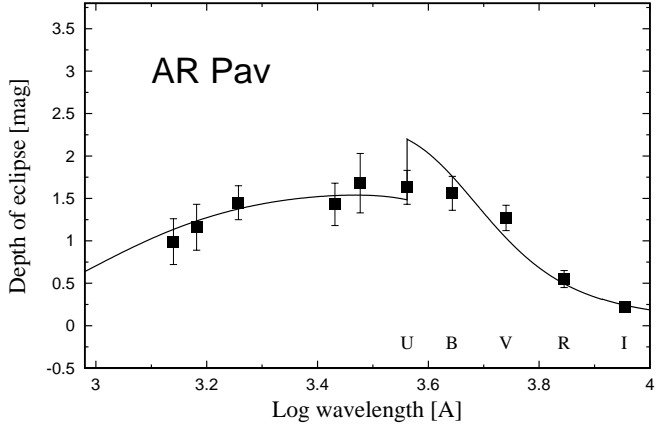}
 \caption[ ]{
Wavelength-dependent depth of eclipses measured in the dereddened 
UV/optical/near-IR light curves (full boxes). The solid line 
represents calculated eclipse depth according to our model in Fig. 4. 
} 
\label{7}
\end{figure}
%
%
\begin{figure}
\centering  
 \includegraphics[angle=-90,width=9cm]{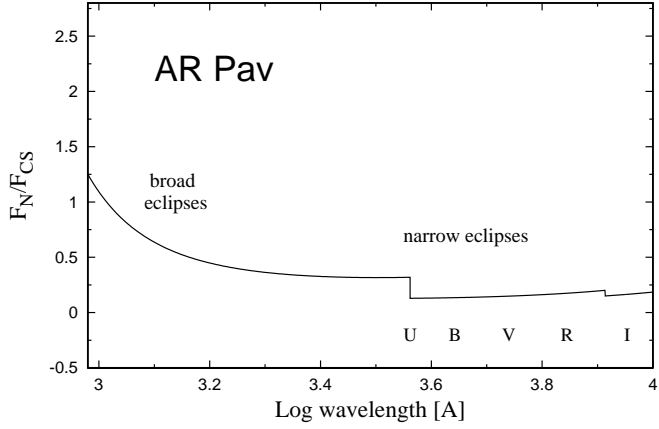}
 \caption[ ]{
The ratio of the hot object fluxes, $F_{\rm N}$ and $F_{\rm CS}$, 
for the solution shown in Fig. 4. This characterizes the broadening 
of the eclipse profile (see text).
}
\label{8}
\end{figure} 

\subsubsection{Width of eclipses}

Observations show a very broad eclipse profile in the far-UV
\citep[$\Delta\varphi\gtrsim 0.2-0.3$ of the orbital phase,][]{schild01},
while in the optical/near-IR the profile is characterized by
a narrow core ($\Delta\varphi\lesssim 0.1$) with shallow wings
\citep{andrews,mal82}. 
The hot object in \astrobj{AR\,Pav} consists of basically two 
components -- the extended nebula and the central star with 
an accretion disk. Let $f_{\rm N}$ and $f_{\rm CS}$ be the fluxes 
emitted by a surface element of the projected nebula and the central 
source, respectively, to the sky. The observed eclipse profile
then can be characterized by the ratio $f_{\rm N}/f_{\rm CS}$.
The case of $f_{\rm N}/f_{\rm CS} \approx 1$ makes the eclipse
profile very broad, because the extended nebula contributes significant
emission with respect to that of the small central source. On the other
hand, $f_{\rm N}/f_{\rm CS} \ll 1$ results in a narrow eclipse core
determined by the central source, accompanied with shallow wings as
the nebula is faint \citep[see also Fig. 4 of][]{sk01b}.
In our approximation and for a spherical nebula, it is simple
to find that
\begin{equation}
 f_{\rm N}/f_{\rm CS} \propto F_{\rm N}/F_{\rm CS},
\end{equation}
where $F_{\rm N}$ and $F_{\rm CS}$ are integrated fluxes of the nebula
and the central source at a given wavelength. Figure 4 shows that 
the ratio $F_{\rm N}/F_{\rm CS}$ is a function of the wavelength,
which thus represents the primary cause of the observed 
wavelength-dependent eclipse width. Therefore, we plotted this 
ratio in Fig. 8 according to modeled components of radiation shown 
in Fig. 4. In the effect of relation (14), Fig. 8 shows that 
\begin{equation}
(f_{\rm N}/f_{\rm CS})_{\rm uv} > (f_{\rm N}/f_{\rm CS})_{\rm opt},
\end{equation}
which answers the question why the eclipses are broader in 
the ultraviolet than in the optical/near-IR region. A more 
rigorous approach, in which the parameter $f_{\rm N}/f_{\rm CS}$
is treated as a function of the radial distance from the central 
object, would give a more accurate information about the nebula 
structure. However, this analysis is beyond the scope of this paper. 

\section{Conclusions}

The results of this study may be summarized as follows: 

(i) 
Based on the ultraviolet IUE low-resolution spectra and the 
broad-band optical and infrared $UBVRI$ and $JHKL$ photometry, 
we re-constructed the SED in the \astrobj{AR\,Pav} 
continuum between 0.12 and 3.4\,$\mu$m. By this way we revealed 
a very high luminosity of the hot object in the binary,
$L_{\rm h} \sim 2\,200\,(d/4.9\,{\rm kpc})^{2}\,L_{\odot}$. 

(ii) 
We failed to fit the UV/optical continuum of the hot object by 
a standard ionization model of symbiotic binaries. This suggests 
that the configuration of \astrobj{AR\,Pav} differs significantly from 
the typical configuration of symbiotic binaries during their 
quiescent phases. 

(iii)
The UV/optical/IR SED can be fitted by a simple blackbody accretion 
disk embedded in an extended hot corona 
($T_{\rm e} = 40\,000\,\pm$5\,000\,K). 
The standard thin disk model requires a high accretion rate 
$\dot M_{\rm acc} > 2 \times 10^{-4}\,M_{\odot}\,yr^{-1}$
onto the central star with a radius $R_{\rm acc} \gtrsim 2\,R_{\odot}$. 
Any accretion process at 
$\dot M_{\rm acc} > 1.7 \times 10^{-5}\,M_{\odot}\,yr^{-1}$
(Sect. 5.1) for $M_{\rm acc} = 0.75 - 1.0\,M_{\odot}$ limits 
the accretor radius to $R_{\rm acc} > 0.1\,R_{\odot}$ 
to balance the observed accretion luminosity. This result suggests 
that the accreting star in \astrobj{AR\,Pav} is non-degenerate. 

(iv)
Te basic configuration of the hot object -- the accretion disk embedded 
in an extended nebula -- allow us to understand the observed 
wavelength-dependent depth and width of the eclipse profile 
(Figs. 7, 8): 
The eclipse {\em width} results from that the hot object consists of 
the two geometrically very different components of radiation (N + AD), 
the ratio of which is a function of the wavelength. 
The eclipse {\em depth} is given by a different ratio of 
the uneclipsed (N+G) to the total light throughout the spectrum. 

(v)
Models with the accretion disk and its BL ($L_{\rm BL} \sim L_{\rm AD}$) 
failed to fit the far-UV SED. However, the presence of a very hot 
region in the system is indirectly indicated through the strong nebular 
emission. This suggests that the hotter inner parts of the disk with 
the BL are permanently occulted by the disk material, when viewing 
the system (nearly) edge-on. 
%

\ack
This research has been supported by the Slovak Academy of Science 
under a grant No. 1157. The author is grateful to Dr. D. Chochol 
for commenting on the manuscript, Dr. H.M. Schmid for his critical 
reading of an earlier draft of this paper, and an anonymous referee 
for several helpful comments.

\end{document}